\documentclass[namedreferences]{SolarPhysics}
\usepackage[optionalrh]{spr-sola-addons} 
\usepackage{graphicx}        
\usepackage{color}           
\usepackage{url}             

\newcommand{\pder}[2]{ \frac{\partial #1}{\partial #2} }
\newcommand{\ppder}[2]{ \frac{\partial^2 #1}{\partial #2^2}}
\newcommand{\pppder}[2]{ \frac{\partial^3 #1}{\partial #2^3}}

\newcommand{\aap}{    {\it Astron. Astrophys.}}

\newcommand{\aapr}{   {\it Astron. Astrophys. Rev.}}

\newcommand{\apj}{    {\it Astrophys. J.}}
\newcommand{\apjl}{   {\it Astrophys. J. Lett.}}
\newcommand{\apjs}{   {\it Astrophys. J. Suppl.}}

\newcommand{\pasj}{   {\it Pub. Astron. Soc. Japan}}

\newcommand{\solphys}{{\it Solar Phys.}}

\begin{document}

\begin{article}

\begin{opening}

\title{Interpretation of Solar Magnetic Field Strength Observations}

\author{R.~K.~\surname{Ulrich}$^{1}$\sep
        L.~\surname{Bertello}$^{1}$\sep
        J.~E.~\surname{Boyden}$^{1}$\sep
        L.~\surname{Webster}$^{1}$      
       }
\runningauthor{Ulrich et al.}
\runningtitle{Magnetic Field Scale Factor}

   \institute{$^{1}$ Department of Physics and Astronomy, University of California at Los Angeles
                     email: \url{ulrich@astro.ucla.edu}\\ 
             }

\begin{abstract}
This study based on longitudinal Zeeman effect magnetograms and spectral line scans investigates the dependence of solar surface magnetic fields on the spectral line used and the way the line is sampled in order to estimate the magnetic flux emerging above the solar atmosphere and penetrating to the corona from magnetograms of the Mt.\ Wilson 150-foot tower synoptic program (MWO). 
We have compared the synoptic program $\lambda5250$\AA\ line of Fe$\,$I to the line of Fe$\,$I at $\lambda5233$\AA\ since this latter line has a broad shape with a profile that is nearly linear over a large portion of its wings. 
The present study uses five pairs of sampling points on the $\lambda5233$\AA\ line.
Line profile observations show that the determination of the field strength from the Stokes $V$ parameter or from line bisectors in the circularly polarized line profiles lead to similar dependencies on the spectral sampling of the lines with the bisector method being the less sensitive.  
We recommend adoption of the field determined with the line bisector method as the best estimate of the emergent photospheric flux and further recommend the use of a sampling point as close to the line core as is practical.  
The combination of the line profile measurements and the cross-correlation of fields measured simultaneously with $\lambda5250$\AA\ and $\lambda5233$\AA\ yields a formula for the scale factor $\delta^{-1}$ that multiplies the MWO synoptic magnetic fields.  
 Using $\rho$ as the center-to-limb angle (CLA), a fit to this scale factor is $\delta^{-1}=4.15-2.82\sin^2(\rho)$. Previously $\delta^{-1}=4.5-2.5\sin^2(\rho)$ had been used.  
The new calibration shows that magnetic fields measured by the MDI system on the SOHO spacecraft are equal to  $0.619\pm0.018$ times the true value at a center-to-limb position 30$^\circ$.
\inlinecite{2003SoPh..213..213B} found this factor to be $0.64\pm0.013$ based on a comparison the the Advanced Stokes Polarimeter.
\end{abstract}
\end{opening}

\section{Introduction}
The sun's 22-year cycle of activity is most clearly seen in solar surface magnetic fields (see for example the review by \inlinecite{2003A&ARv..11..287O} for a good discussion of the solar dynamo and solar magnetic fields).  
Although the Sunspot Number (SSN) is the most readily available indicator of the state of the solar cycle, the magnetic fields over the whole surface provide a more complete measure of the dynamo process.  
A variety of questions arise in using the surface magnetic fields to study the solar cycle -- which fields are most important, the weak general field or the strong field and associated sunspots, how do the field strengths change over various time scales from instabilities with changes in seconds or less to trends lasting many decades.  
The connection between photospheric magnetic fields and magnetic fields near earth or at interplanetary spacecraft requires knowledge of the field strength at the solar surface. 
Is the overall strength of the magnetic field stationary or does it have any multi-decade trends (cf: \inlinecite{2002JGRA.107jSSH16A})?  
These surface fields are also needed to calculate the strength measured by interplanetary spacecraft such as {\it Ulysses} \cite{1999ApJ...520..871G}. 
The topology of the field in the solar surface regions introduces additional uncertainty since the field of one polarity lower in the atmosphere may be canceled by the opposite polarity before it can emerge into the corona and heliosphere \cite{2007ApJ...671..936A}.  
Questions such as these involving magnetic field strength can only be addressed when we are confident in the quantitative interpretation of the measured quantity.

The 150-foot solar tower telescope on Mt.\ Wilson has been dedicated to the synoptic measurement of solar surface magnetic fields using a Babcock magnetograph observing the Fe I spectral line at $\lambda5250$\AA.  Although other spectral lines and observing techniques offer advantages over these choices, the uniformity of the data set extending back to 1967 is a strong reason to leave the observing system as nearly unaltered as possible.  The application of these data to modeling the heliospheric magnetic field then depends on our ability to understand how to interpret the magnetograms.  An early step toward this goal was carried out by \inlinecite{1972SoPh...22..402H} and subsequently extended by \inlinecite{1992ASPC...26..265U} and 
\inlinecite{2002ApJS..139..259U} using a cross-correlation of fields observed simultaneously from two or more pairs of spectral samples.  A compelling feature of the scatter diagrams from these simultaneous measurements is the linearity between the magnetic fields obtained from the different spectral samplings.  Each spectral sample measures a different altitude in the solar atmosphere and is influenced by different thermal properties of the atmosphere at that altitude.  If the altitude dependent part of the relationship were also to depend on the field strength at each point on the solar surface, the scatter diagram would not be confined to a straight-line band passing through the zero-zero point.  The observed straight line distributions found on the scatter diagram establish that variable filling factor is the only cause for varying field strength, all flux tubes being essentially identical.  The fact that the slope is not unity can be used to study the altitude dependence of the flux tubes.  

Throughout this paper we will use the term ``magnetic field strength'' and usually mean the apparent field strength which is the product of a flux tube field strength and a filling factor.  This quantity is actually the magnetic flux per unit area and is the quantity we wish to measure.  The extensive literature devoted to the determination of the flux tube fields and the filling factors is of interest to us and is referenced where appropriate below.  However, questions of flux tube physics are of secondary importance to our objectives in this paper. Where the flux tube field strength is discussed, it is made explicit.  Otherwise, the term magnetic field strength should be read to mean the magnetic flux per unit area as measured in the observed pixels.

The essential step in our study of magnetic field strength is the choice of a comparison line.  Our objective is to learn the relationship between measured circular polarization of a relatively large pixel (12 arcseconds squared or 20 arcseconds squared) and the magnetic flux in that area.  We are less concerned with questions of the structure of magnetic flux tubes or the strength of the field within the flux tubes -- we only seek to learn the best way to estimate the product of the filling factor and the magnetic field in the flux tubes.  In addition, we need to be able to observe the line simultaneously with $\lambda5250$\AA.
For this purpose the line at $\lambda5233$\AA\ is ideally suited due to its large width and depth and due to the good linearity of its line wings.  For a related objective of determining the flux tube magnetic field strength the comparison line chosen by \inlinecite{1973SoPh...32...41S} was $\lambda5247$\AA\ to take advantage of the fact that the two lines have differing $g_{\rm eff}$ values but very similar thermal response and line formation properties.  However in both this paper and the subsequent study by \inlinecite{1978A&A....70..789F} the reference to $\lambda5233$\AA\
 was retained with the assumption that magnetic fluxes from this line are obtained directly from the polarization factors and the line profile.  An interesting citation to a private communication by Livingston in this last paper suggests that the magnetic fields derived from $\lambda5233$\AA\ depend on position within this line.  We verify below that the deduced magnetic field strength for any particular area on the sun's surface that depends on the spectral sampling of $\lambda5233$.
Our previous results based on $\lambda5233$\AA\ yielded a correction factor of 4.5 to be applied as a multiplier for the magnetic fields obtained from $\lambda5250$\AA\ whereas \inlinecite{1973SoPh...32...41S} found only a factor of two.  Our correction factor is larger due partly to the fact that the shifted Zeeman components at $\lambda5250$\AA\ are shifted beyond the sampling pass-band of the MWO Babcock magnetograph.  We refer to the reduced apparent field due to the large line shift as the ``classical saturation effect'' since it depends only on the actual line profile, the strength of the flux tube magnetic fields and the spectral resolution of the observing system. 
\section{Magnetogram Observations}
\label{scat}
As a first step in establishing a more complete understanding of the scale factors relating different observing methodologies we have carried out a new series of observations utilizing simultaneous magnetic field measurements with six different spectral configurations. The MWO system integrates all spectral channels continuously with all integrator registers being emptied sequentially at a rate of 40 Hz.  Due to this feature of the system image motion and smearing affects all spectral samples identically so that the spatial average of spatial inhomogenieties is also identical.  Thus ratios of magnetic fields calculated from different sampling pairs yield relative magnetic fields from identical solar surface structures.  Starting with the concept that the magnetic field from $\lambda5233$\AA\ indicates the true magnetic flux, the ratio $\delta^{-1}=B_{\lambda5233\hbox{\scriptsize\AA}}/B_{\lambda5250\hbox{\scriptsize\AA}}$ should be the scale factor that converts the MWO synoptic program magnetic fields to the appropriate value emerging from the photospheric layers.  For the actual observations we need to specify the spectral sampling as well as the spectral line.  The study by \inlinecite{1992ASPC...26..265U} used $\lambda5250\hbox{\AA}\pm39$m\AA\ and $\lambda5233\hbox{\AA}\pm45$m\AA\ to yield the scale factor of 4.5.  Contrary to the assumptions of \inlinecite{1972SoPh...27..330F}, \inlinecite{1973SoPh...32...41S} and \inlinecite{1978A&A....70..789F} we find that the field measured by $\lambda5233$\AA\ depends on the spectral sampling so that we must generalize the notation.  We do this by defining a new quantity $\eta^{a,n}_{b,m}$ which relates the magnetic field for the spectral line at wavelength $a$ having spectral sampling offset by $n$ to the magnetic field for the spectral line at wavelength $b$ having spectral sampling of $m$ with $a$ and $b$ in \AA ngstroms and $n$ and $m$ in m\AA.  Thus we write:
$$B_{a,n}=\eta^{a,n}_{b,m}B_{b,m}$$
so that the previously defined $\delta^{-1}$ is:
$$\delta^{-1}=\eta^{5233,45}_{5250,39}\ .$$
The fact that the magnetic field derived from $\lambda5233$\AA\ depends on the sampling separation is not a surprise in view of the complexity of the atmospheric flux tube structure but it does present us with the additional task of determining which portion of this line is most appropriate as an indicator of the emergent magnetic flux.  

\begin{figure}
\begin{center}
\parbox{4.8in}{
\begin{center}
\resizebox{4.8in}{!}{\includegraphics{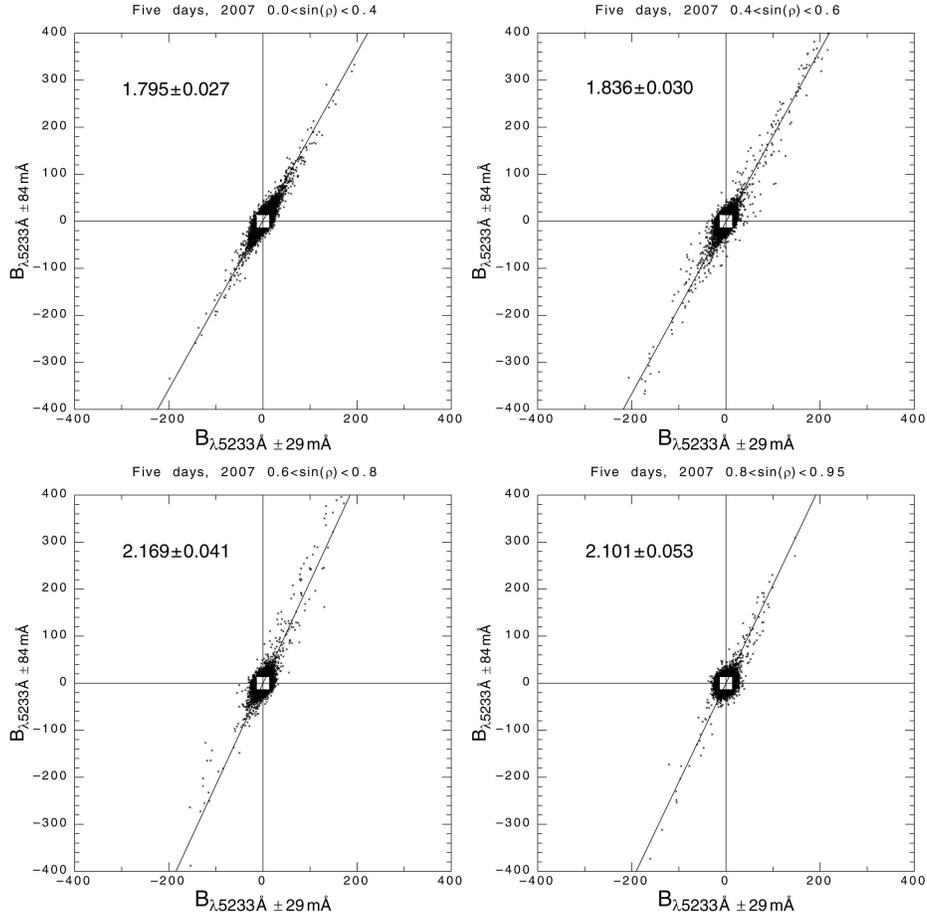}}
\end{center}
}
\end{center} 
\caption{This figure shows the relationship between magnetic fields determined from spectral samples at $\lambda 5233\hbox{\AA}\pm 29$m\AA\ and at $\lambda 5233\hbox{\AA}\pm84$m\AA.  Each point on the figure shown as the very small \textsc{x} represents a single pixel observed once. The range in center-to-limb angle $\rho$ for each figure is shown above each figure. The light at the sampling wavelengths is recorded over identical time and space limits so that the difference in determined field strengths is a result of differing physical conditions at the two different heights of formation.  The portion of the figure near the 0,0 origin is without points in order to improve the manageability of the plot by decreasing the number of points plotted.  The slope of the line relating the two determined field strengths is indicated in the upper left section of each plot.  This quantity $\eta^{5233,84}_{5233,29}$ multiplies the field determined from $\lambda 5233\hbox{\AA}\pm 29$m\AA\ in order to convert that field into the value that would be determined from the combination at $\lambda 5233\hbox{\AA}\pm84$m\AA.  The scatter of the points is due to photon statistics as is discussed in the text.}
\label{f1}
\end{figure}
\begin{figure}
\begin{center}
\parbox{4.8in}{
\begin{center}
\resizebox{4.8in}{!}{\includegraphics{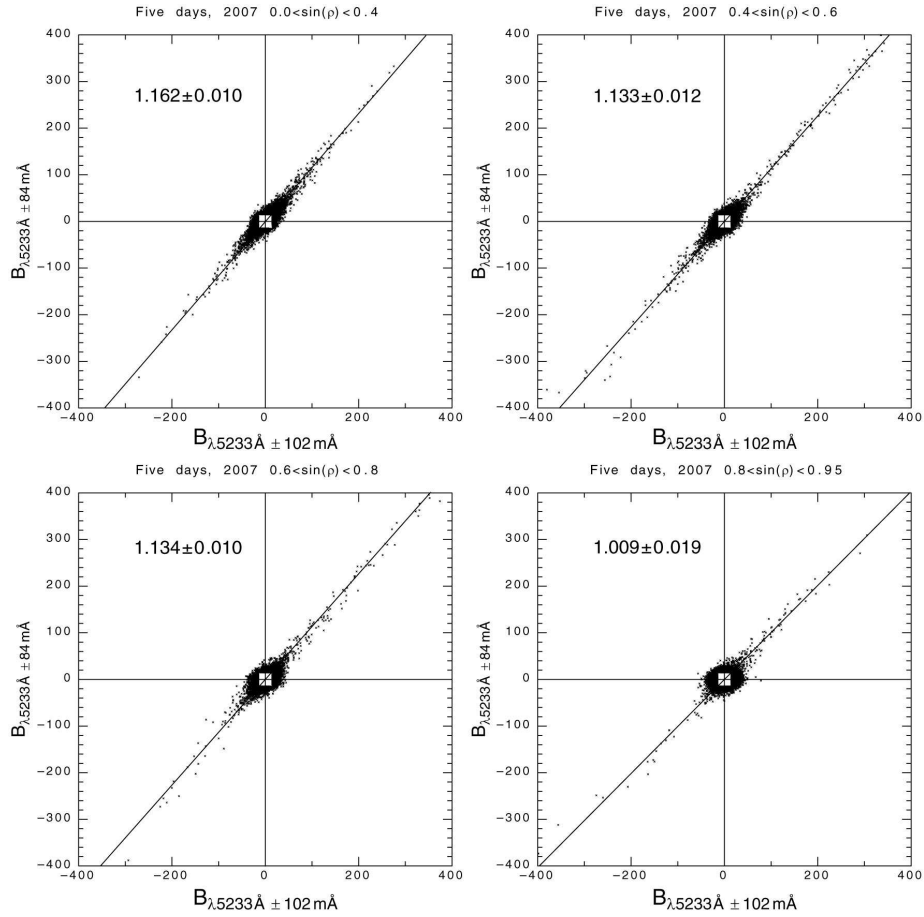}}
\end{center}
}
\end{center}
\caption{ This figure is in the same format as Figure \ref{f1} but shows the relationship between magnetic fields determined from spectral samples at $\lambda 5233\hbox{\AA}\pm 102$m\AA\ and at $\lambda 5233\hbox{\AA}\pm84$m\AA.  The tightness of the relationship is noteworthy.}\label{f2}

\end{figure}
%

\begin{figure}
\begin{center}
\parbox{4.8in}{
\begin{center}
\resizebox{4.8in}{!}{\includegraphics{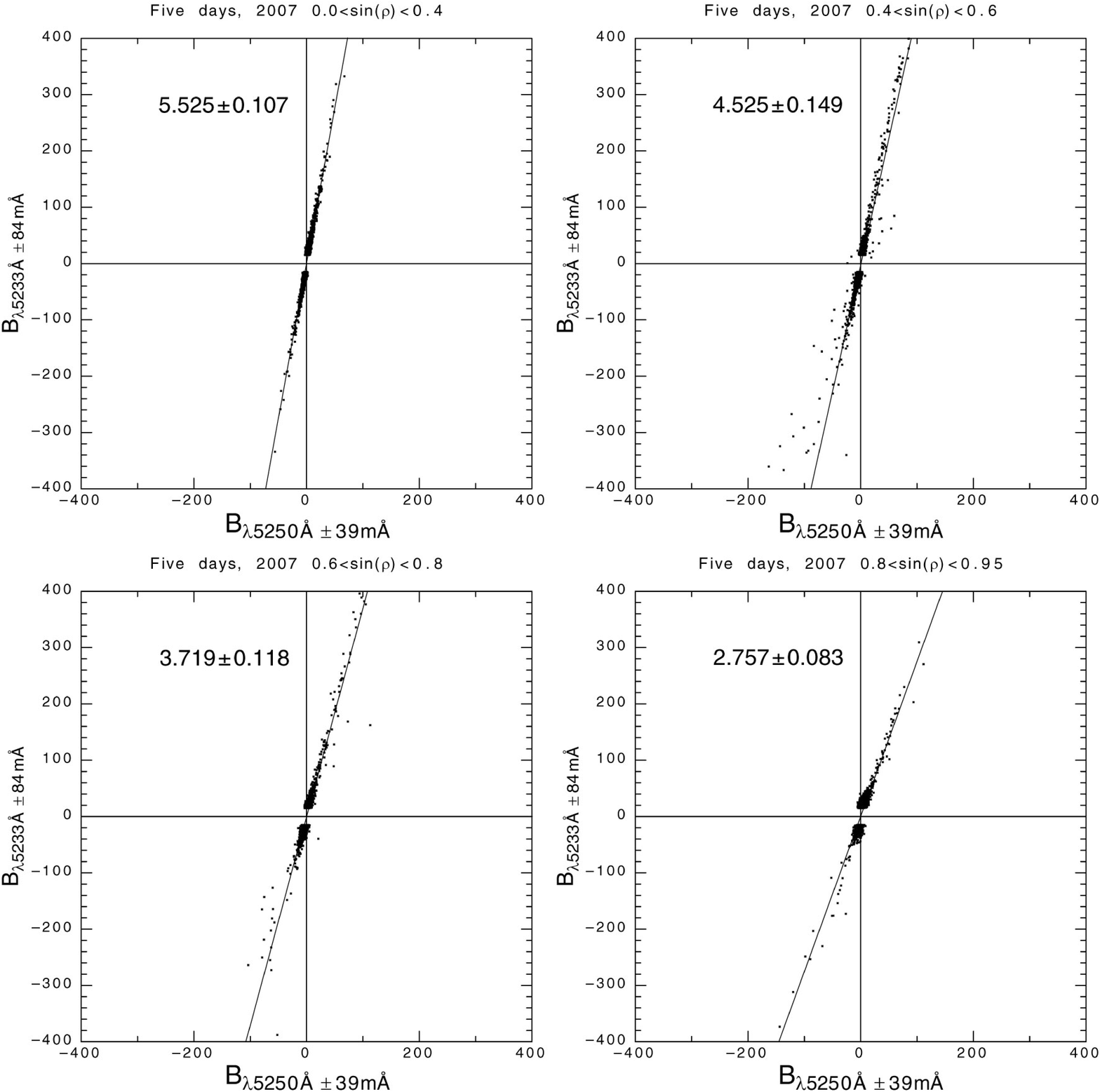}}
\end{center}
}
\end{center}
\caption{ This figure is in the same format as Figure \ref{f1} but shows the relationship between magnetic fields determined from spectral samples at $\lambda 5250\hbox{\AA}\pm 39$m\AA\ and at $\lambda 5233\hbox{\AA}\pm84$m\AA.}\label{f3}

\end{figure}

During the interval 18 April 2007 to 8 May 2007 a set of five special magnetograms were obtained using the 24-channel system described by \inlinecite{2002ApJS..139..259U} with the blue, 10-channel fiberoptic reformattor centered on $\lambda5233$\AA\ instead of the $\lambda5896$\AA\ line of NaD.  This configuration provided us with five line pairs having separations of $\pm9, \pm29, \pm84, \pm102\hbox{ and } \pm177$m\AA.  The observations in 1992 were made using a different spectrograph configuration wherein the spectral line sampling was at $\pm45$m\AA.  The sampling of $\lambda5250$\AA\ was left as in the synoptic program at $\pm39$m\AA.  
During the scanning process the light entering the spectrograph is unfiltered.  The dispersed light passes through narrow-band blocking filters where it falls on all spectral sampling pickups simultaneously.  The output currents from all photomultiplier tubes are summed individually and simultaneously into a set of registers.  We refer to the sequence from the entry window of a single spectral sample to the summing register as a channel.  The final digitization numbers for all channels for each pixel then originate from identical portions of the solar surface even though the sampled area is not precisely defined by the entrance aperture due to distortions of the solar image by the Earth's atmosphere.  We compute magnetic fields from each spectral sampling pair and compare the derived fields in a scatter-diagram format.  Since we expect the interrelationships to depend on position within the solar image, we restrict the scatter diagrams to pixels having a specified range in center-to-limb angle $\rho$.

\begin{table}
\caption{The correlation slopes $\eta^{5233,84}_{5233,m}$ and their errors as a function of $m$ and $\sin(\rho)$.}
\label{t0}
\begin{tabular}{lr@{$\pm$}lr@{$\pm$}lr@{$\pm$}lr@{$\pm$}l}
\hline
$m$&\multicolumn{2}{c}{\hspace*{-.08in}$0<\sin(\rho)<0.4$\hspace*{0.05in}}&\multicolumn{2}{c}{\hspace*{-.08in}$0.4<\sin(\rho)<0.6$\hspace*{0.05in}}&
\multicolumn{2}{c}{\hspace*{-.08in}$0.6<\sin(\rho)<0.8$\hspace*{0.05in}}&\multicolumn{2}{c}{\hspace*{-.08in}$0.8<\sin(\rho)<0.95$}\\
\hline
9&1.665&0.067&1.887&0.043&2.262&0.077&2.042&0.105\\
29&1.795&0.027&1.836&0.030&2.169&0.041&2.101&0.053\\
102&1.162&0.010&1.133&0.012&1.134&0.010&1.009&0.019\\
177&2.694&0.094&2.727&0.085&2.706&0.075&2.607&0.206\\
\hline
\end{tabular}
\end{table}

The comparisons were done between the various samplings and the one at $\lambda5233\hbox{\AA}\pm84$m\AA.  The corresponding scatter diagrams for three of the five cases are shown in Figures \ref{f1} to \ref{f3}. The values of $\eta^{5233,84}_{5233,m}$ are given in Table~\ref{t0} for all five cases.  The adoption of $n=84$ is somewhat arbitrary but was done because the intensity slope in the line is large and nearly linear at this point.  Furthermore, this part of the line is well correlated with the two adjacent sampling points: the Pearson's $r$ correlation coefficient has values of 0.50, 0.90, 0.90 and 0.50 for $m=9,\, 29,\, 102 \hbox{ and } 177$ and is 0.96 for $b,\, m = 5250,\, 39$.  The coefficients and their uncertainties have been computed using the methods described by \inlinecite{1990ApJ...364..104I} and \inlinecite{1992ApJ...397...55F} wherein neither of the two quantities is treated as dependent or independent.  The multiplicative factors to convert the magnetic fields to the $\lambda5233\hbox{\AA}\pm84$m\AA\ scale are given in Table~\ref{t0} and deviate from each other.  This clearly indicates that $\lambda5233\hbox{\AA}$ does not provide a single reference field strength.  The deviation of the scale factor for $\lambda5233\hbox{\AA}\pm29$m\AA\ from unity is particularly significant in that these field measurements are well correlated with those from $\lambda5233\hbox{\AA}\pm84$m\AA\ in spite of the fact that the line wing profile is straight and neither is close enough to the line core for the flux tube field to shift the Zeeman component past the sampling point.    In the next section we develop an algebraic model for the saturation factor and then in section~\ref{profinterp} apply this model to observed profiles of the $\lambda5233$\AA\ line to verify the above conclusion.

\begin{figure}
\begin{center}
\parbox{4.8in}{\parbox{2.4in}{\resizebox{2.4in}{!}{\includegraphics{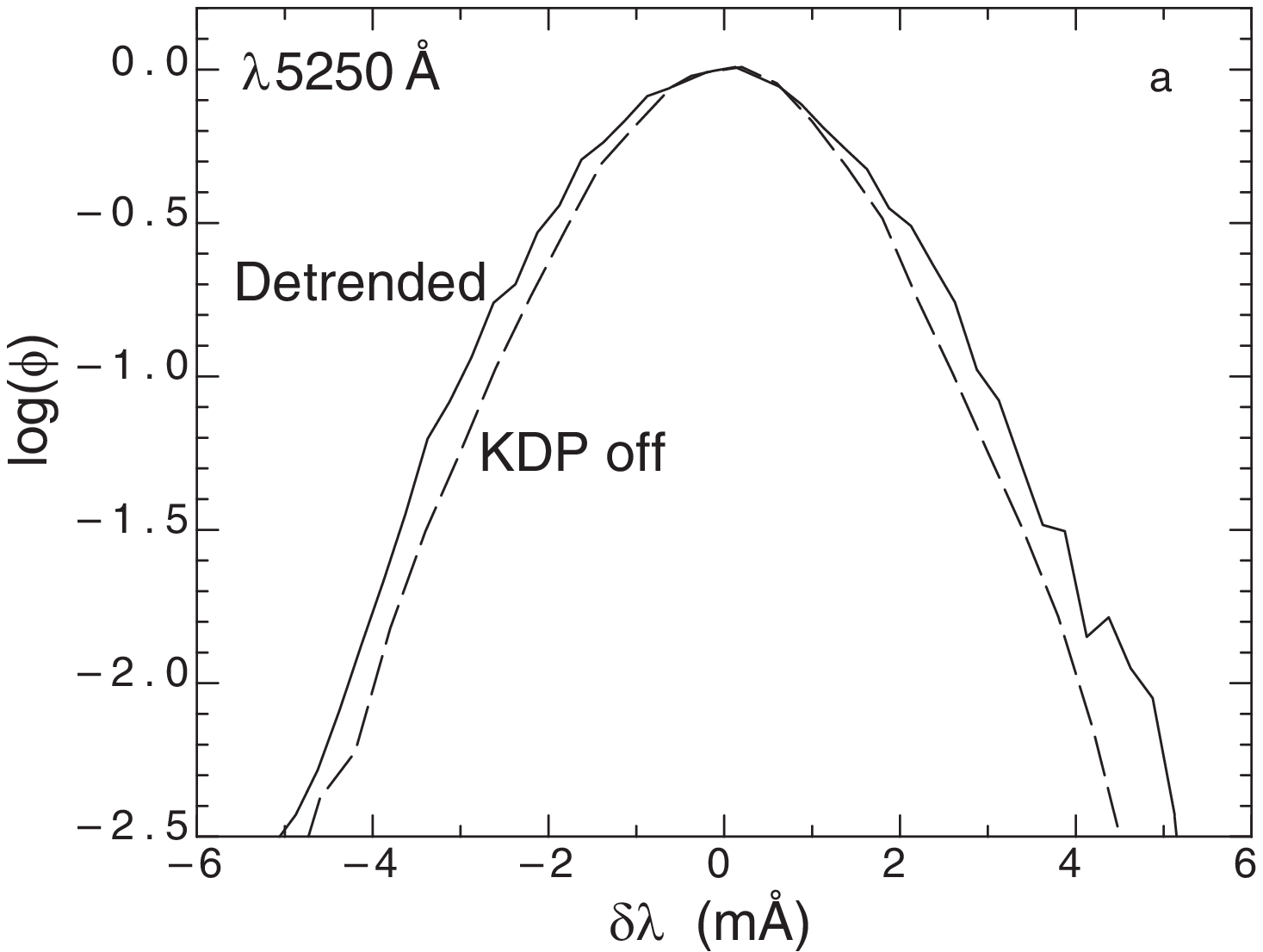}}}\hfill\parbox{2.4in}{\resizebox{2.4in}{!}{\includegraphics{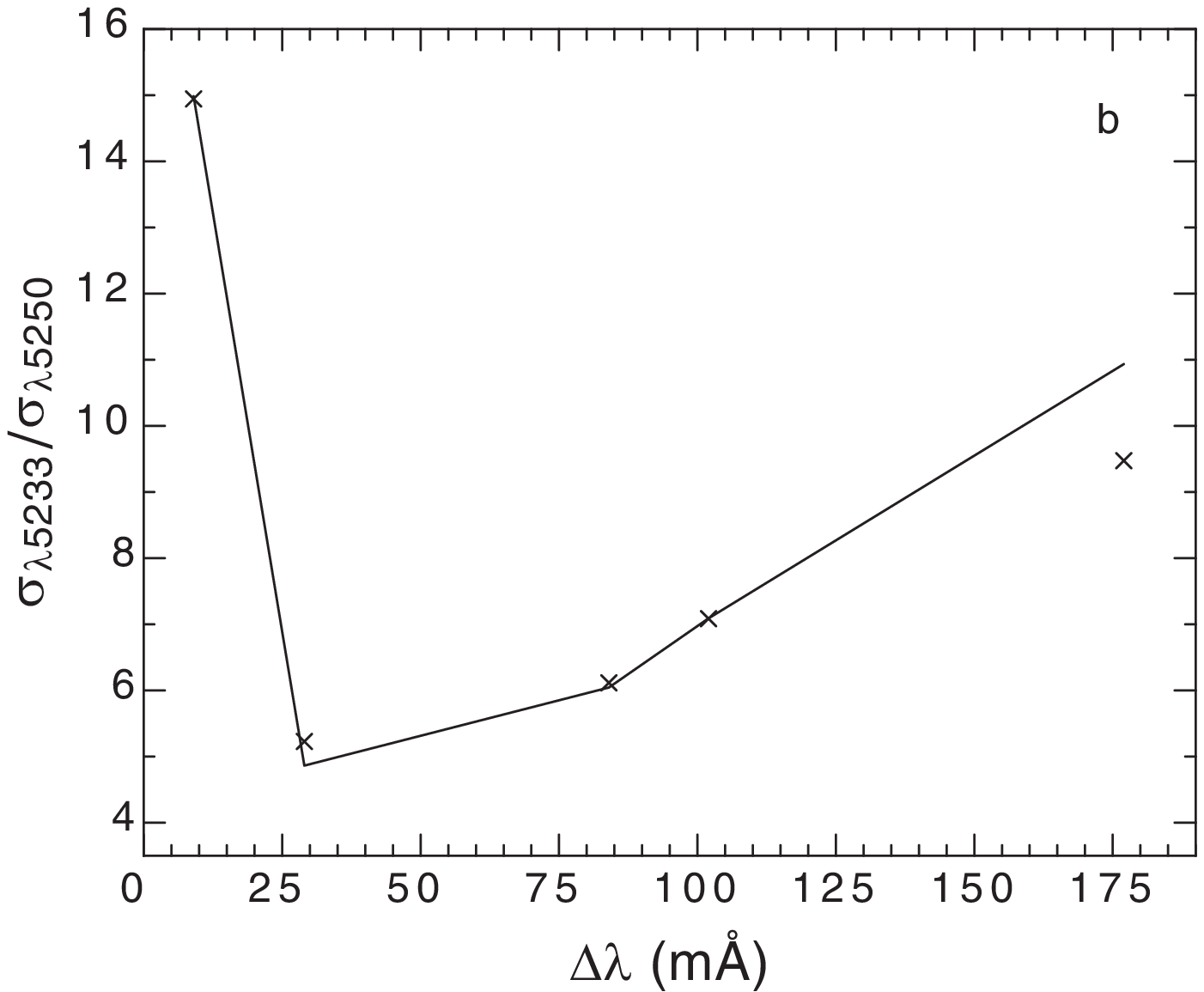}}}}
\caption{Part a of this figure shows the spread of points on the scatter diagram (labeled Detrended) compared to the spread found from a standard magnetogram observed with the polarization modulating KDP analyzer inactive (labeled KDP off).  Part b of this figure shows the ratio of the gaussian width of the detrended scatter diagrams for the five parts of the $\lambda5233$\AA\ line shown as the symbols compared to a model that assumes that the variation in measured field is due to photon noise shown as the solid line.}\label{f4}
\end{center}
\end{figure}

One issue that arises later in section~\ref{obs} is the cause of the spread in the scatter diagrams shown in Figures~\ref{f1} to \ref{f3}.  Two possible causes are instrumental noise mostly due to photon statistics or an intrinsic deviation from a perfectly linear relationship between the different pairs of spectral sampling points.  We distinguish between these two explanations by examining distribution functions $\phi(x)$ where $\phi(x)$ is the number of pixels having $x$ between $x$ and $x+\Delta x$ and we use a normalization where $\phi(0)=1$.  For $x$ we use either $x=B-B_{\rm fit}$ or $x=B_{\rm KDP~off}$.  For $B_{\lambda5250\hbox{\AA}\pm39\hbox{m\AA}}$ the value of $B_{\rm fit}$ for each point is derived from $B_{\lambda5233\hbox{\AA}\pm84\hbox{m\AA}}$ using the trend line in Figure~\ref{f3}.  For $B_{\rm KDP~off}$ we use an observation from the regular synoptic magnetogram program where for calibration and control purposes we periodically measure the magnetic field while the KDP modulator is not active.  For the KDP~off observation, the apparent magnetic field comes only from the system noise.  Figure~\ref{f4}a shows this comparison.  Clearly most of the scatter in Figure~\ref{f3} comes from photon or system noise.

A second part of the identification of the cause for the spread on the scatter diagrams is an estimation of the photon noise caused spread as a function of the spectral sampling separation $\Delta\lambda$.  The deviation of $V$ is the combined result of the deviations of the intensity measured for opposing states of the KDP analyzer with the form of the dependence being the same for all $V$'s.  Each intensity has a fractional error that is inversely proportional to the number of photons collected.  These intensity errors are converted into errors in $V$ and subsequently into errors in $B$.  The factor converting $V$ errors into $B$ errors is inversely proportional to $g(dI/d\lambda)/I$.  The slopes and $g$ values are known so the model for the relative errors only depends on the estimation of the relative number of photons measured for each spectral channel.  All channels for $\lambda5233\hbox{\AA}$ use one set of fiber-optic image reformattors while the channel for $\lambda5250\hbox{\AA}$ uses an independent reformattor.  The relative number of photons measured for all $\lambda5233\hbox{\AA}$ channels is just proportional to the relative line intensity at each spectral sampling point.  Between these channels and the one for $\lambda5250\hbox{\AA}$ there is an uncertain relative transmission parameter.  If we pick this single parameter to give the best fit then we can compare the spreads for all five spectral samples at $\lambda5233\hbox{\AA}$ to each other.  This comparison is shown in Figure~\ref{f4}b.  Clearly the model that the detrended magnetic field spreads is a consequence of photon noise is well established.

\section{The classic saturation effect}
\label{cse}
The classical saturation effect comes from an interaction between an observing system and the fact that the observed pixels for the solar surface are much larger than the magnetized portions of the solar atmosphere.  Although the consequences of solar inhomogenieties have been intensively studied for many years \cite{1993SSRv...63....1S,2006RPPh...69..563S}, the shape of the $\lambda5233\hbox{\AA}$ line has led to the assumption that this line does not suffer from the classical saturation effect.  In view of the fact that we have found the observed magnetic fields derived from this line to depend on the spectral sampling, we feel it is important to quantitatively evaluate the classical saturation effect and definitively rule it out as a contributor to the spectral sampling dependence we find. 

The inhomogeneous character of the solar atmosphere means that the detected radiation is a combination with a fraction $f$ from the flux tubes having a large Zeeman shift and a possibly different line shape and a fraction $1-f$ coming from unmagnetized plasma with no Zeeman shift and a profile that may be closer to that from the quiet sun.  The line shift amount $\Delta\lambda_H$ is related to the magnetic field strength by the standard formula:
$$
\Delta\lambda_H=4.67\times10^{-13}g_{\rm eff}\lambda^2B
$$
where $\lambda$ is in \AA\ and $B$ is in gauss.  (There is a notation conflict between this standard usage of $\Delta\lambda_H$and the quantity $\Delta\lambda$ introduced in section \ref{bisector} below.  The usage with subscript $H$ is restricted to this section and the later usages do not include this subscript.)  Due to the non-linear shape of the line profile and the possible large shift in the flux tubes, it is not appropriate to calculate the observed change in the line intensity from $f\Delta\lambda_H (\partial{I}/\partial{\lambda})$.  In addition, the magnetogram observations do not measure the spectral line position for the two states of circular polarization since the movable stage does not translate at the rate at which the polarizations are switched.  The stage servo drives the stage position towards the point at which the intensity averaged between the polarization states is equal for the spectral sampling ports on the blue and red wings of the line.  The position of the polarized line is then inferred from the Stokes $V$ parameters at blue and red wing points on the line where the $I$ parameters are equal.  Thus the operation of the system is then based on formulae originally discussed by \inlinecite{1984A&A...131..333S} with $V$ being the difference in intensity for line profiles shifted by the Zeeman splitting amount.  These can be represented
using Taylor expansions for $I(\lambda\pm\Delta\lambda_H)$ up to third order.
In the presence of an inhomogeneous atmosphere structure where the magnetized regions occupy a fraction $f$ of the total area the formula for the $V$ parameter can be written:
$$
V = f\left(\pder{I}{\lambda}\Delta\lambda_H + \frac{1}{6}\pppder{I}{\lambda}(\Delta\lambda_H)^3\right)\ .
$$
Our interest is in the combination $B_{\rm ave}=fB_{\rm FT}$.  The field strength and  corresponding Zeeman shift $\Delta\lambda_{H_{\rm FT}}$ are taken to be consistent with a model external to our analysis and the deduced value of $B_{\rm ave}$ is then a function only of $f$.  The third order relationship of interest for $B_{\rm ave}$ can be written
$$
B_{\rm ave}=\frac{V}{4.67\times10^{-13}g_{\rm eff}\lambda^2(\partial I/\partial\lambda)}\beta(\lambda)
$$
where he quantity in front of $\beta(\lambda)$ is just the standard formula for the magnetic field according to the weak field approximation.  We denote the field derived from this term as $B_{\rm Stokes}$ and use this uncorrected quantity as one of the primary ways of estimating the field strength.  The quantity $\beta(\lambda)$ is a saturation correction factor that depends on the line shape near the sampled point and the strength of the flux tube magnetic field which is given by
$$
\beta(\lambda)=1 - \frac{\Delta\lambda_{H_{\rm FT}}^2}{6}V^{-1}\ppder{V}{\lambda}\ .
$$
Due to the difficulty in calculation of high order derivatives from observed data, we have replaced the first derivative of $I$ by the Stokes $V$ parameter.
A similar replacement of the line slope by a function of $V$ was recommended by \inlinecite{1993SSRv...63....1S} for the study of flux tube field strength using the magnetic line ratio method.  In our formula $\beta(\lambda)$ depends only on the choice of flux tube magnetic field strength and not on the filling factor.  The final formula based on $V$ also has the advantage that it comes from the magnetized gas whereas the formula based on $I$ includes radiation from both magnetized and unmagnetized gas. Due to the difficulty of taking the second derivatives, we only use this formula to estimate the size of the classical saturation factor and show that this effect cannot be responsible for the differences in magnetic field found from different parts of the line profile (see section \ref{sat} below).

The above equations indicate the relationship between the circular polarization of the radiation and the magnetic field strength.  However, the intensity emerging from individual flux tube which should be used in these equations is not available for observational determination under conditions of spatial resolution that prevail at the 150-foot tower telescope.  The calibration observations carried out routinely prior to the magnetogram scans consist of spectral line scans in a quiet part of the solar atmosphere to yield $I_{\rm quiet}(\lambda)$.
In principle the saturation factor $\beta(\lambda)$ should be calculated from the intensity $I_{\rm FT}(\lambda)$ that emerges from a spatially unresolved flux tube.  For reasons discussed by \inlinecite{2001A&A...369..646F} and as apparent from line profiles observed by our system and shown in Figure~\ref{f6} below, $I_{\rm FT}(\lambda)\ne I_{\rm quiet}(\lambda)$.  Despite the line profile sensitivity to magnetic field, the tight correlations shown in Figures \ref{f1} and \ref{f2} demonstrate that the line shape effects are consistent over a large fraction of the solar surface so that the correction and filling factor effects in one area of the surface can be applied to other similar areas. 

\section{Relationship between line profiles and magnetograms}

\subsection{Context}

At each point of observation we wish to have an algorithm that returns a single value for the magnetic field strength even though we find a dependency of the deduced field strength on the profile sampling of the $\lambda5233$\AA\ line. We verify  below the expectation that the classical saturation effect does not explain this dependence so we must use other considerations to make this recommendation.  As is well established \cite{1993SSRv...63....1S,2000RvGeo..38....1L,2006RPPh...69..563S} the atmosphere consists of a weakly magnetized gas having one thermal stratification threaded by a varying density of highly magnetized flux tubes each of which has a different thermal stratification, the spectral line samples all of these structures with relative contribution factors that depend on position within the line.  

The strength of the magnetic field within the flux tube is generally thought to vary with altitude \cite{1994IAUS..154..407S} and the magnetic flux per pixel (the product of the field strength, the filling factor and the pixel area) may or may not be constant through the atmosphere since it is possible for field lines to close within the atmospheric zone.  The radiation traversing the atmosphere containing these structures can be influenced by vertical and horizontal gradients that differ from the mean atmosphere.  For example differently shifted Zeeman components of the line might produce an effect similar to microturbulence for an emerging ray path over which the magnetic field strength is decreasing.  These considerations lead to the fact that the shifted line profiles for the two states of circular polarization are generally not the same as the line profile in an unmagnetized region shifted by a constant amount and further that they are not shifted by the same amount in all their parts.

The line profile from an unmagnetized region gives us the residual intensity as a function of wavelength offset from line centre.  An intensity difference such as that between the two states of circular polarization is then used to infer a line shift from the unmagnetized profile from an inverse where wavelength is considerd to be a function of residual intensity. Although the shift of the flux tube line is large, the dilution with non-magnetized radiation makes the intensity difference also small so that the line shift can only be interpreted using the local non-magnetized line slope instead of the fully non-linear line shape information.  The difference between this interpretation and one that would be possible if we had adequate spatial resolution to isolate each flux tube is what we refer to as the classical saturation effect.

\subsection{Heights of formation}

Our task is to interpret the line profiles in terms of the flux-tube model and relate shifts in the line to a field strength appropriate to each part of the line and appropriate for each position on the solar disk.  As a first step we have carried out a height of formation calculation for both $\lambda5233$\AA\ and $\lambda5250$\AA\ using the methods described by \inlinecite{1977A&A....54..227C}.  We then deduced the standard optical depth from the atmosphere model of \inlinecite{1981ApJS...45..635V} and inferred the magnetic field strength from the models of \inlinecite{2001A&A...369..646F} at the corresponding optical depth.  The field strengths are given for our working points and center-to-limb positions in Table~\ref{t1}.  These tables also give as $\Delta\lambda_H$ the line shift amount appropriate for the Zeeman splitting due to the field strengths at the working points.

\begin{table}
\caption{The heights of formation, flux-tube field strengths and Zeeman shift amounts.}
\label{t1}
\begin{center}
\parbox{4.8in}{
\begin{center}
\resizebox{4.8in}{!}{\includegraphics{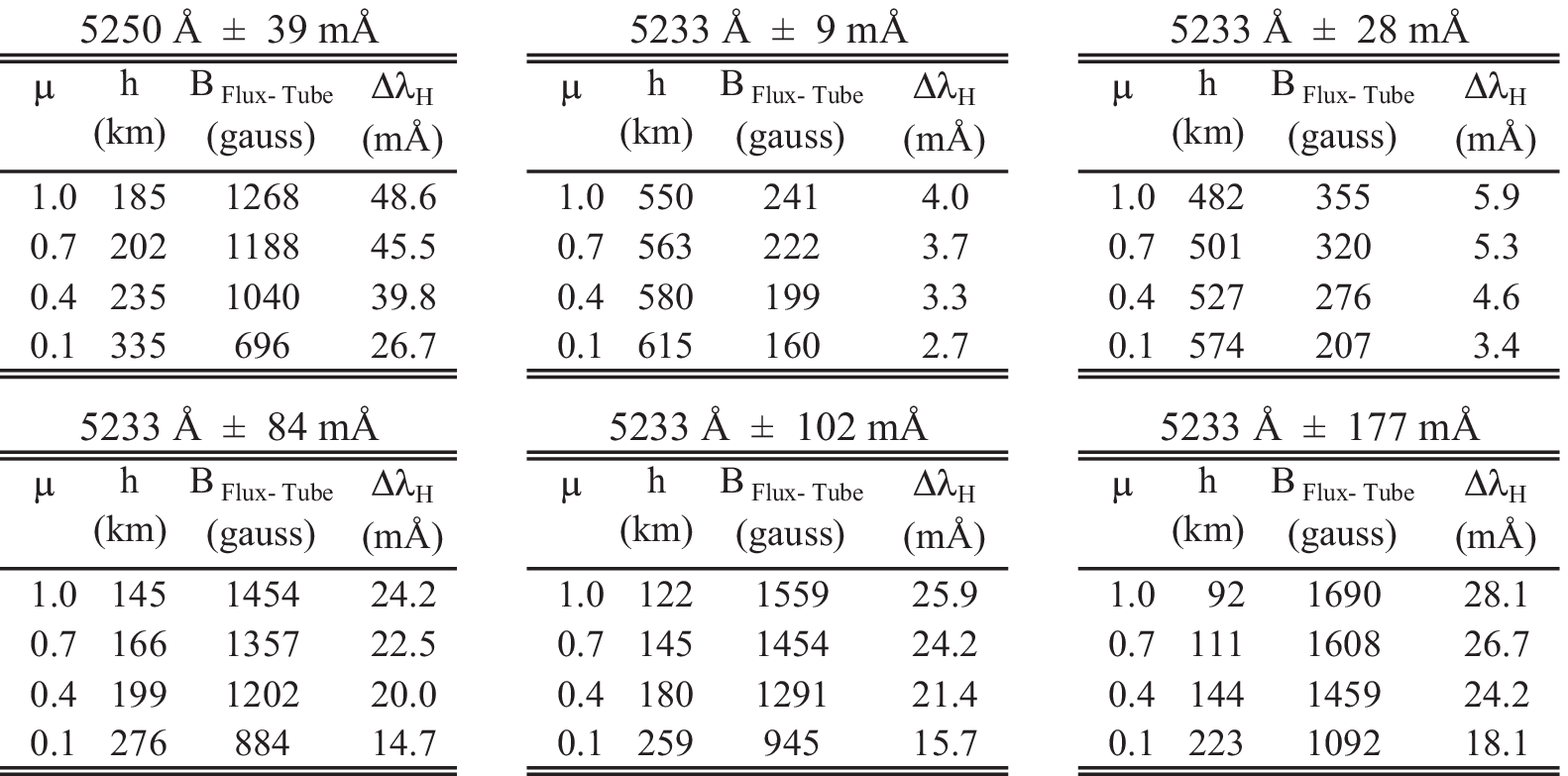}}
\end{center}
}
\end{center}
\end{table}

\subsection{A line bisector definition of the magnetic field}\label{bisector}

Solar surface magnetograms provided by the system at Mt.\ Wilson's 150-foot tower telescope are deduced from the longitudinal Zeeman effect starting with the measured intensity difference between two or more samples of a spectral line in the two states of circular polarization.  As described by \inlinecite{1983SoPh...87..195H} a double difference among the four measured intensities yields an indication of the opposing line shifts of the two states of circular polarization.  The Babcock magnetograph includes a mechanism to track the Doppler shift of the observed line and attempts to maintain equal average intensity for the spectral samples on the blue and red wings of the line.  The mid-point between the blue and red equal intensity wavelengths is commonly referred to as a spectral line bisector point.  The bisector points depend on position within the spectral line which can be defined either in terms of the ratio of the intensity to the continuum intensity (often called the residual intensity) or in the wavelength separation between the sampled blue and red wings of the line.  Because the Babcock magnetograph defines a fixed wavelength separation, it is natural for us to adopt the bisector half-width $\Delta\lambda$ as the parameter to define the position within the spectral line. (Recall that here $\Delta\lambda$ without subscript $H$ is not the Zeeman splitting used in section \ref{cse}).   This choice has the additional advantage that it is relatively unaffected by differences in the central residual intensity that we find to be common in comparing magnetized and unmagnetized regions.  

\subsection{Magnetic fields from spectropolarimetry modelling}

Another way to interpret the polarimetry of spectral lines starts with the forward theory such as has been formulated by \inlinecite{1956PASJ....8..108U} and \inlinecite{1973SoPh...31..299L}.  The polarized monochromatic intensity is then found from a solution to the transfer equation including the fact that the atoms in the magnetic field respond differently to the different states of light polarization.  The resulting system of equations depends on the detailed stratification of the magnetic field and the physical state of the plasma -- a highly complex combination that can only be managed with the help of approximations and simplifications.  The fact that even the highest spatial resolution observations include contributions from magnetized and non-magnetized solar plasma means that a potentially very large number of free parameters is present in the task.  It is especially unfortunate that spectra from individual flux tubes are not available so that the line shape from an individual flux tube must be recovered from the model rather than being observed directly.  

One form of solution is based on the use a Milne-Eddington approximation \cite{1987ApJ...322..473S} and does not treat gradients in magnetic field strength.  By including two or more spectral lines and invoking flux tube models with structural detail, \inlinecite{2001A&A...369..646F} have been able to estimate the thermal, dynamic and magnetic stratification of the flux tubes.  Typically studies following this approach take the Stokes $V$ polarization parameter as a function of wavelength near the spectral line as the primary constraining data.  Works by \inlinecite{1999A&A...349..941S} and \inlinecite{2005A&A...442.1059K} show that a range of Stokes $V$ profiles is encountered for high spatial and temporal resolution observations.  As a consequence of this variability, it is likely that the underlying physical structure including the height dependence of the magnetic field is also variable.  However, our interest is in the use of magnetograms for the purpose of establishing an inner boundary condition for global-scale magnetic models of the heliosphere.  For this purpose, it is appropriate to use coarse spatial and temporal resolution which is less impacted by phenomena at or near the solar granulation scale.  Indeed, the scatter diagrams of Figures \ref{f1} to \ref{f3} show that there is a tight correlation between the nominal magnetic fields derived from different parts of $\lambda5233\hbox{\AA}$ and between $\lambda5233\hbox{\AA}$ and $\lambda5250\hbox{\AA}$.  

\subsection{Comparison to Advanced Stokes Polarimetry fields through MDI as an intermediary}\label{profandmag}

Although most studies using the techniques based on multi-line Stokes polarimetry and inversions have concentrated on determination of the thermal and magnetic structure of the flux tubes, the work by \inlinecite{2003SoPh..213..213B} determined the relationship between the magnetic flux per unit area obtained with the Advanced Stokes Polarimeter (ASP) and the apparent field strength measured by the MDI magnetogram program.  They found that the MDI fields are reduced to $64\pm1.3$\%\ of the fields deduced from the ASP for the non-sunspot portions of an area centered around Active Region 8218 at S20 W22 on 13 May, 1998 with a center-to-limb angle of 30$^\circ$.  For a similar position on the solar disk, the study by \inlinecite{2005ApJS..156..295T} found that the MDI fields are reduced to 55\%\ of the magnetic fields at $\lambda5250$\AA\ measured by the MWO synoptic program after correction to the scale previously defined by $\lambda5233$\AA.  This comparison shows that the fields we previously deduced using the $\lambda5233$\AA\ scale are 16\%\ greater than those deduced using the multi-line spectropolarimetry/inversion method.

\section{Line Profiles}

\subsection{Observations}
\label{obs}

\begin{table}
\caption{Line scan observations of $\lambda5233\hbox{\AA}$}
\label{t2}
\parbox{3.2in}{
Time is UT, Latitude, Central Meridian Distance (CMD) and Center to Limb Angle (CLA) are in degrees, the absolute value of the magnetic field $|B|$ is in gauss and has been determined according to the symmetrized Stokes $V$ method below evaluated at $\delta\lambda=84$m\AA.}
\begin{tabular}{crrrrr}
\hline
\multicolumn{1}{c}{Set No.}&\multicolumn{1}{c}{Date\ Time}&\multicolumn{1}{c}{Lat.}&\multicolumn{1}{c}{CMD}&\multicolumn{1}{c}{$|B|$}&\multicolumn{1}{c}{CLA}\\
\hline
022&2007/07/13\ 18:39&-4.93&5.56&558\phantom{.0}&10.7\\
023&2007/07/13\ 18:52&-2.89&-0.35&15.8&7.2\\
025&2008/03/31\ 19:01&-11.32&-54.49&24.5&54.0\\
026&2008/03/31\ 19:24&-2.43&-53.17&5.2&53.1\\
027&2008/03/31\ 19:56&-10.83&-70.45&  14.3& 69.60\\
028&2008/04/02\ 18:08&-5.26&-80.60&  19.2& 80.02\\
029&2008/04/02\ 18:30&0.43&-75.32&  2.3& 75.49\\
030&2008/04/05\ 18:05&-8.70&-66.45&  68.6& 65.86\\
031&2008/04/05\ 18:17&-8.69&-66.58& 63.2& 65.96\\
039&2008/06/14\ 18:21&-9.43&2.48& 103\phantom{.0}& 10.82\\
040&2008/06/14\ 18:51&3.19&-0.78& 16.2 & 2.21\\
041&2008/06/15\ 22:03&-9.72&-12.48& 8.4& 16.55 \\
042&2008/06/15\ 22:25&-7.87&-15.04& 132\phantom{.0}& 17.59\\
043&2008/06/15\ 22:47&-0.01&-15.05& 12.6& 15.13\\
044&2008/06/21\ 18:41&-1.93&-5.16& 161\phantom{.0}&  6.44\\
045&2008/06/21\ 19:04&-1.92&3.34& 325\phantom{.0}&  5.18\\
046&2008/06/21\ 19:25&-1.93&6.33& 3.8 & 7.49\\
047&2008/06/22\ 18:40&-1.99&-19.17& 314\phantom{.0}& 19.58\\
048&2008/06/22\ 19:05&-0.65&-10.05& 174\phantom{.0}& 10.45\\
049&2008/06/22\ 19:29&1.22&-0.60&  2.7&  1.02\\
050&2008/06/26\ 19:06&-1.20&-62.56&  72.5& 62.64\\
051&2008/06/26\ 19:28&-1.89&-71.48&  53.8& 71.56\\
052&2008/06/26\ 19:50&-1.33&-57.32&  2.7& 57.41\\
\hline
\end{tabular}
\end{table}
\begin{figure}
\begin{center}
\parbox{4.8in}{
\begin{center}
\resizebox{4.8in}{!}{\includegraphics{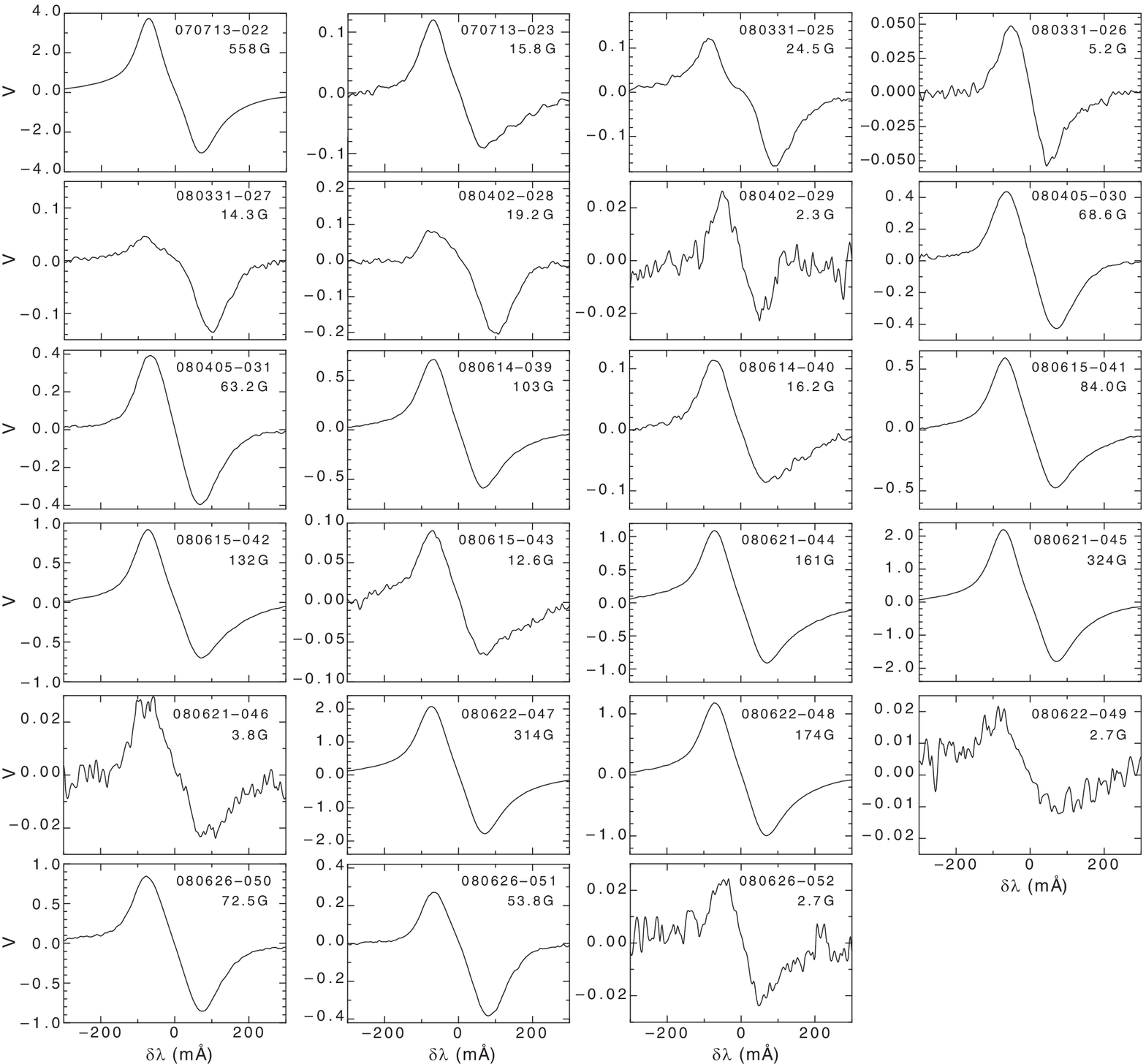}}
\end{center}
}
\end{center}
\caption{ This figure shows the Stokes $V$ profiles for the 23 sets of line profile scans.  The scale of $V$ is in percent relative to the continuum $I$.  Each figure is labeled by the date of observation and sequence number allowing other parameters to be found in Table~\ref{t2}.  The signal to noise of the polarimetry in these Stokes $V$ profiles is approximately a factor of 10 greater than previously published cases and thus these measurements carry the analysis to field levels that are weaker than previously studied.  In fact most of the variance in the $V$ parameter as a function of wavelength and consequently also in the derived field strength comes from fringing effects introduced by our blocking filter.  The estimated error in the derived field strength for the weaker cases is 0.1 gauss.  The magnetic field as determined from the Stokes $V$ parameter at $\pm84$m\AA\ is given on the figures as well as being included in Table~\ref{t2}.}\label{f5a}

\end{figure}

To investigate the properties of the Stokes $V$ profiles and deduced magnetic field strengths based on such profiles, 23 line scan sets were obtained over a period of several months during the declining phase of solar cycle 23.  We use observations obtained between June and August 2007.  At this stage of the solar cycle, the number of active regions available for study is limited.  The observing system is that described by \inlinecite{1991SoPh..135..211U} and utilized the 10-channel fiber-optic reformatter normally used for magnetograms in the $\lambda5896\hbox{\AA}$ line of sodium (see \inlinecite{2002ApJS..139..259U} for the details of the spectral sampling).  With this system we obtain spectral line profiles from a square selected by our entrance aperture which has sides of 12 arc-sec (other apertures are available but were not used in this study). This entrance aperture tracks solar rotation using the nominal differential rotation rate for the position observed.  At the final spectrograph focus the spectrum is sampled by one of four sets of fiber-optic image reformatters that are carried on two independently movable stages.  A small portion of the solar spectrum including the target spectral line is scanned by moving its stage alternately redward and blueward over the solar spectrum repeatedly for a total duration of 20 minutes with each scan requiring 20 seconds. The second stage is left stationary and its output is used to remove sky transparency fluctuations. The length of each scan is 1\AA\ and the spectral output from each fiber is sampled at an interval of 5.31m\AA\ (the spectral resolution is about 30m\AA).  The temporal resolution was not used and all scans were summed to a single profile for each spectral pickup.  Finally the ten profiles were shifted by the known fiber-optic separation so that all are superposed for a single final profile.  The offset between the fiber-optic inputs means that the portion of the spectrum scanned is also offset.  The overall scan length was large enough that all ten inputs covered the range of interest for comparison to the magnetogram scatter diagram results.  The circular polarization modulation was retained to permit the calculation of a single Stokes $V$ profile.  Following the convention described by \inlinecite{2000RvGeo..38....1L} the blue shifted component is left circularly polarized in a positive magnetic field region.  Although we retain knowledge of the polarity of each set of line profiles, we do not consider the sign of the field to be important in this application and present the Stokes $V$ parameter and the derived magnetic fields as if all polarities are positive.  Table \ref{t2} gives the parameters for these scans including the nominal magnetic field strength based on the line bisectors for a bisector width $\Delta\lambda$ of $\pm84$m\AA.   The line scan set numbers given in Table~\ref{t2} are not sequential because a number of line scan sets for $\lambda5250$\AA\ were also obtained during this time frame and these observations will be discussed in a future publication.

The Stokes $V$ profiles from the observations summarized in Table~\ref{t2} are shown in Figure~\ref{f5a}. The scale of the Stokes $V$ plots is based on a continuum $I$ of unity.  The discussion in the following sections emphasizes the set of line scans from July 13, 2007 since that region was more highly magnetized than was the case for subsequent observations.  The apparent noise for the low field strength Stokes $V$ profiles comes from a combination of factors including photon noise and the formation of interference fringes by filters, windows and other plane-parallel optics in the system.  We have mitigated these residual fringes by tilting the blocking filter but the effect is not eliminated by this technique.  We estimate the uncertainty in the Stokes $V$ profile over the spectral range needed for the field determination is $V_{\rm error}\approx 5.0\times10^{-6}$.  The line slope is $3\times10^{-3}$(m\AA)$^{-1}$ at the points $\pm84$m\AA\ from line center so that the Zeeman shift error due to this noise is about $1.7\times10^{-3}$m\AA\ for an estimated uncertainty in the field strength of 0.1 gauss as observed by the channel at $\lambda5233\hbox{\AA}\pm84\hbox{m\AA}$.  We can also estimate the photon noise error in the deduced magnetic field using measurements from the normal synoptic program magnetograms. The integration time for each magnetogram pixel is 0.127 seconds while the integration time on each pixel in the spectral scans is very similar at 0.1 seconds.  The final profile information is the result of combining scans from all 10 fiber-optic reformattor channels and from all 60 scans.  Thus the relative photon noise for each line profile point compared to the synoptic program magnetograms should be reduced by a factor of $(600/1.27)^{1/2} = 22$.  The line scanning process provides four times as many spectral points compared to the magnetogram leading to a further reduction of 2 in the expected photon noise.  According to the discussion of section~\ref{scat}, the gaussian width from the magnetograph distribution function is 4 gauss for the above spectral channel so that again we arrive at an estimate of 0.1 gauss for the spectral scans.  For observations with the synoptic program channel at $\lambda5250$\AA\ the equivalent error would be 0.025 gauss.  These errors are substantially lower than has been previously reported and allow us to study the magnetic fields in regions that would normally be classified as unmagnetized.

\subsection{Interpretation}
\label{profinterp}
For this section, we exclusively use the profile sequence number 022 of July 13, 2007 to illustrate different ways of interpreting the profiles. 
Figure \ref{f6} compares the quiet and plage profiles.  The two states of circular polarization are combined for the low magnetization profile but separated for the magnetized case.  Because we are not concerned here with Doppler shifts, the low magnetization line has been shifted so its bisector position coincides with the average bisector position for the two magnetized components at a bisector half-width $\Delta\lambda$ of 75 m\AA.  This parameter measures the product of the filling factor and the flux tube field strength and represents the magnetic flux emerging from the pixel. Not only do the two circularly polarized line profiles differ significantly from the quiet sun profile, they also differ from each other in such a way as to yield a magnetic flux which depends on position within the profile.  The strength of this line and its high altitude of formation at line core make it difficult to attribute this effect to typical flux tube physics where thermal effects can produce line weakening or strengthening relative to the surrounding quiet atmosphere.  Clearly then this spectral line does not define a single magnetic flux and our problem becomes that of deciding how to interpret the observations in such a way as to find the best estimate of the magnetic flux emerging from the solar atmosphere.  

\subsubsection{Magnetic field estimated from the line bisector.} 
\begin{figure}
\begin{center}
\parbox{3.8in}{
\resizebox{3.8in}{!}{\includegraphics{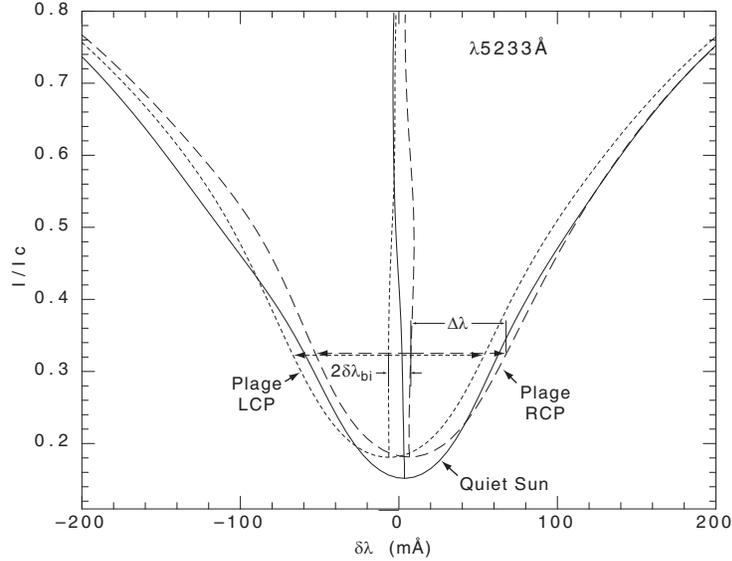}}}
\end{center}
\caption{This figure compares the line profiles from a quiet sun region to the two circularly polarized components from an active region which was magnetized with a negative polarity field. The quiet sun profile is shown as the solid line while the LCP and RCP polarized profiles are shown as the short dashed and long dashed lines respectively.  For this figure the polarized profiles are normalized to the continuum being unity instead of 0.5 as is required for the calculation of the Stokes parameters.  Also shown with the same dashed/undashed indication are the line bisectors for each of the three profiles.  The parameter $\Delta\lambda$ defining the position on the profile through the half-width value is shown for the two circularly polarized profiles.  Finally this figure also shows the parameter $\delta\lambda_{\rm bi}$ which gives the Zeeman splitting of the profile at the selected value of $\Delta\lambda$.  The magnetic field at this position on the profile is found from $\delta\lambda_{\rm bi}$ using the standard formula for the Zeeman effect and for this active region according to Table \ref{t2} the field strength was 558 gauss.}
\label{f6}
\end{figure}

\begin{figure}
\begin{center}
\parbox{3in}{
\resizebox{3in}{!}{\includegraphics{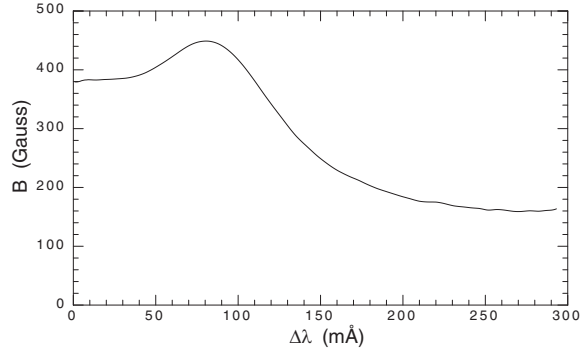}}
\caption{The magnetic field calculated from the difference in wavelength of the line bisectors of the two states of circular polarization.  The position within the line of the bisector is defined by the bisector half width $\Delta\lambda$.  Note that this field differs from that given in Table~\ref{t2} wherein the fields from the Stokes $V$ parameter are given.}
\label{f7}}
\end{center}

\end{figure}

The Stokes $V$ profile is often used as the basis for this estimate.  As an alternate way of thinking about the different parts of the spectral line it is worth considering how a Babcock magnetograph might work.  If we could simultaneously track the spectral line in both states of circular polarization, the Babcock magnetograph would evaluate the magnetic field on the basis of central wavelengths where the red and blue wing intensities are equal in each state of polarization.  The wavelength difference between these two balance points would then define a magnetic field from the spectrally resolved line profiles.  In general, a line bisector can be considered as either a function of the intensity relative to the continuum intensity or as a function of the wavelength separation between the equal intensity points.  In view of the dependence of the line core intensity on the strength of the magnetic field, we find it advantageous to use bisector half-width $\Delta\lambda$ as the defining parameter.  Corresponding bisectors for the two states of circular polarization at a fixed but arbitrary value of $\Delta\lambda$ are shown in Figure \ref{f6}.  The separation between the two central wavelengths at this $\Delta\lambda$ is indicated as $\delta\lambda_{\hbox{bi}}$ from which a magnetic flux per unit area $B_{\rm bi}(\Delta\lambda)$ can be deduced. For the observation shown in Figure \ref{f6} we show the derived bisector field strength as a function of $\Delta\lambda$ in figure \ref{f7}.  


\begin{figure}
\begin{center}
\parbox{3in}{
\resizebox{3in}{!}{\includegraphics{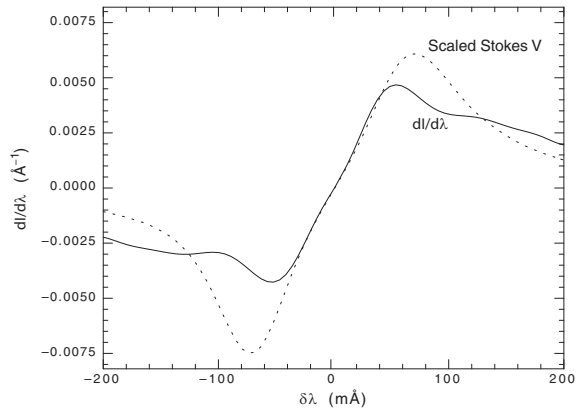}}
\caption{A comparison of the slope of the spectral line in the non-magnetized comparison region to the Stokes $V$ parameter from the nearby magnetized region. The Stokes $V$ parameter has been scaled to have the same shape near the line center.  In the positively magnetized region the observed Stokes $V$ parameter reached a maximum of 3.7\%\ at -72 m\AA\ and a minimum of -3.1\%\ at +70 m\AA.}
\label{f8}}
\end{center}

\end{figure}

\begin{figure}
\begin{center}
\parbox{3in}{
\resizebox{3in}{!}{\includegraphics{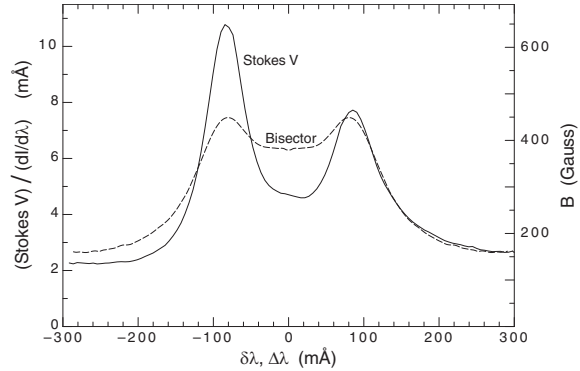}}
\caption{This figure shows the ratio of the Stokes $V$ parameter divided by the spectral line slope.  The left scale gives the result in units of m\AA\ while the right scale gives the inferred magnetic field strength in gauss.  In this figure, the asymmetric $V$ profile is used.  Also shown is the magnetic field deduced from the line bisector separation using the bisector width $\Delta\lambda$ as being equal to the distance from line center $\delta\lambda$.  This function is plotted for both $+\Delta\lambda$ and $-\Delta\lambda$ since it is symmetric as derived.}
\label{f9}}
\end{center}

\end{figure}

\begin{figure}
\begin{center}
\parbox{3in}{
\resizebox{3in}{!}{\includegraphics{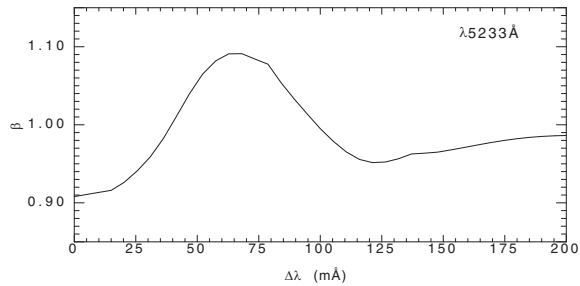}}
\caption{This figure shows the classical saturation factor $\beta$ for the line $\lambda5233$.  The flux tube field was taken to be 1500 gauss as is consistent with points in the wings of the line.  Near line core where the field strength is about 400 gauss, the value of $\beta$ is much closer to unity.}
\label{f10}}
\end{center}

\end{figure}

\begin{figure}
\begin{center}
\parbox{4.8in}{
\resizebox{4.8in}{!}{\includegraphics{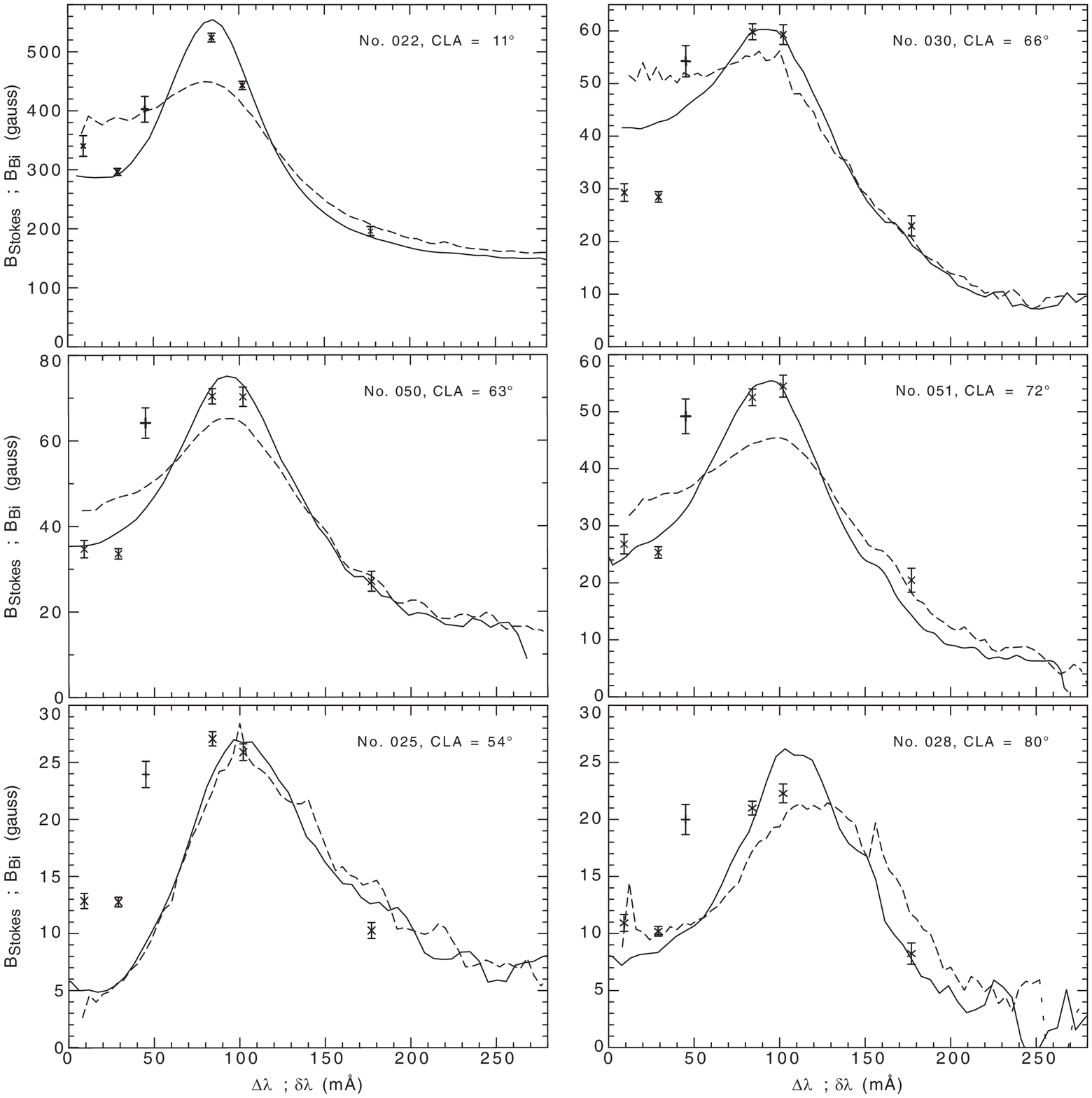}}
\caption{
 This figure shows comparisons between the field strength dependence on position in the spectral lines and the slopes of the correlations given in Table~\ref{t0}.  Each panel corresponds to a line scan set from Table~\ref{t2} as indicated on the panel.
The solid lines are for $B_{\rm Stokes}(\delta\lambda)$ and the dashed lines are for $B_{\rm Bi}(\Delta\lambda)$.  The Stokes $V$ profile from Figure \ref{f9} has been made symmetric by averaging the blue and red wing results since we are applying these results to magnetograph observations where the blue and red wings are averaged by the instrument.  The field strengths in gauss for both the bisector separation and the Stokes $V$ ratio divided by the slope are given.  The relative correlation slopes have been scaled to correspond to the field from the Stokes $V$ ratio. The \protect\raisebox{-2pt}{\protect\rotatebox{45}{\textsc{+}}} symbols are from Table \ref{t1} while the  \protect\raisebox{0pt}{\protect\rotatebox{0}{\textsc{+}}} symbol is taken from the report by \protect\inlinecite{1992ASPC...26..265U}.
}
\label{f11}}
\end{center}

\end{figure}

\subsubsection{Magnetic field estimated from the Stokes $V$ parameter.}
Figure \ref{f8} which gives both the Stokes $V$ parameter and the slope of the quiet sun line profile.  The scale factor for the Stokes $V$ has been adjusted so that the curves agree near the line center.  The magnetic field can be calculated from this scaling factor if we use the linear approximation.  This gives a field strength based on the point-wise slope of the spectral line and the corresponding Stokes $V$ value so that a magnetic field can be deduced from both wings of the spectral line.  We refer to a magnetic field deduced this way as $B_{\rm Stokes}(\delta\lambda)$. These fields along with the values of $B_{\rm Bi}(\Delta\lambda)$ are shown in Figure \ref{f9}. Note for this plot we use $\delta\lambda=\pm\Delta\lambda$. While $B_{\rm Stokes}(\delta\lambda)$ can be calculated for both red and blue wings of the line, $B_{\rm Bi}(\Delta\lambda)$ applies to both wings equally and has been plotted symmetrically. Although the Stokes $V$ parameter gives independent magnetic fields in the blue and red wings of the line, a Babcock magnetograph responds to the average magnetic field from the two line wings.  Consequently, apart from figures \ref{f8} and \ref{f9}, the magnetic field derived from the Stokes $V$ parameter has been made symmetric by averaging the blue and red wing field values.  The comparison between $B_{\rm Stokes}(\delta\lambda)$ and $B_{\rm Bi}(\Delta\lambda)$ indicates that the pointwise slope leads to a smaller deduced field strength than the bisector by 25\%\ near the line core, a larger field in the mid part of the line wing then good agreement in the outer line wings.  We do not have a definitive interpretation of this result but point out that it could be a consequence of microturbulent broadening of the line core in the flux tube.  The influence of variable microturbulent line broadening on the interpretation of spectropolarimetry of $\lambda5250$\AA\ has been discussed recently by \inlinecite{2008ApJ...674..596S}.

\subsubsection{Classical saturation factor estimation.}\label{sat}
Because the line wings are nearly linear over much of the line, the classical saturation effect does not influence the measurement. The classical saturation factor derived from the Stokes $V$ parameter is shown in Figure~\ref{f10}.  In order to illustrate $\beta$ for a case of interest, a flux tube field of 1500 gauss was used for this calculation although the appropriate field strength for the line core is closer to 400 gauss according to Table \ref{t1}.  Even with this higher than appropriate field assumption, the value of $\beta$ deviates from unity by only 10\%.  The $\lambda5233$\AA\ has nine Zeeman components.  As a simple numerical test, the unmagnetized profile was shifted and weighted by these nine amounts and a simulated profile was then produced.  This numerical test gave essentially the same result as we find by shifting the line an amount equal to the weighted mean of the Zeeman factors.  We conclude from this simulation that the saturation factor is not substantially influenced by the presence of the nine sub-components.  These parts of the line may however play a role in the line transfer process in a manner that is not captured by our simple approach; the multiple Zeeman components may produce the equivalent of an excess of microturbulence.  We conclude that the dependence of the deduced field strength on position within the spectral line for $\lambda5233$\AA\ is not consequence of the classical saturation effect even for points near the line core.

 The $\beta$ parameter can also be derived for the $\lambda5250$\AA\ line.  In this case the greater curvature of the line profile leads to $\beta=1.6$ at our working point of $\pm39$m\AA.  The factor is only 35\%\ of the slope of the relationship indicated by figure~\ref{f3} confirming that thermodynamic effects play a dominant role in the modification of the apparent magnetic field strength.

\section{Magnetic fields referenced to $\lambda5233\hbox{\AA}$}
\label{magfield}
At each point on the solar surface we now have a range of estimates for the magnetic field: $\eta^{5233,84}_{b,m}B_{b,m}$ with $b,m = 5233,9,\; 5233,29,\; 5233,102,\; 5233,177$ and $5250,39$, $B_{\rm Bi}(\Delta\lambda)$ and $B_{\rm Stokes}(\delta\lambda)$.
We should choose the estimate which gives the product of the field strength and filling factor appropriate to the pixel area contribution to the magnetic flux above the solar atmosphere.  For this purpose, the portion of the line nearest the line center would appear to be the best choice since this part of the line is highest in the atmosphere.  However, as we show in this section, the choice has to be tempered by the fact that the line core shows an excess range of variability and is vulnerable to line broadening uncertainties from microturbulence.  

For portions of the solar surface outside of sunspots the apparent magnetic field strength is not greater than about 500 gauss.  For Zeeman shifts from fields of this strength the linear approximation is applicable and the field deduced from a magnetograph observation depends on the spectral line slope at the point of the line where the spectrograph sampling is selected.  This means that the magnetogram field is consistent with what we derived above from the ratio of the Stokes $V$ parameter to the local spectral line slope -- i.e.\ the magnetogram magnetic field should be consistent with the Stokes $V$ magnetic field interpretation of the spectral line rather than the bisector interpretation of the spectral line even though the the Babcock magnetograph follows the average spectral line bisector.  Thus, it is reasonable to expect that the observed magnetogram correlations shown in figures~\ref{f1} to \ref{f3} and Table~\ref{t0} between different deduced fields should be consistent with the Stokes $V$ deduced fields.
The line profile observations come from a relatively small number of selected regions on the solar surface that may or may not be representative of a general case.  However, since many of the slopes from different spectral sampling configurations are well correlated, these particular observed profiles must be representative of the general case since otherwise, there would be a wider scatter in the correlation diagrams.  Thus the implied field values from different parts of the line profile can be normalized so that the Stokes $V$ field strengths and bisector field strengths give scale factors that can be used in bringing the different deductions to a common value.  

The 23 sets of line scans have been reduced to yield the three estimates of the field strength indicated above. The value of $B_{5233,84}$ was chosen to provide a set of field strengths roughly similar to the other two estimates.  The value of $\eta^{5233,n}_{5233,84}$ was interpolated in $\sin(\rho)$ to be appropriate to the position of each line scan field strength function. 
Selected line scan results illustrating this comparison are shown in Figure \ref{f11}. The \protect\raisebox{-2.5pt}{\protect\rotatebox{45}{\textbf{+}}}'s indicate the rescaled correlation coefficients obtained from the magnetograms, the solid and dashed lines give the symmetrized fields from Figure \ref{f9}.  In addition to the cases of $\eta^{a,n}_{5233,84}$ from Table \ref{t0} we have included the implied slope at $\lambda5233$\AA$\pm45$m\AA\ by combining the slope in Figure \ref{f3} with the correlations from the observations of 1991.  This point is shown as the \protect\raisebox{0pt}{\protect\rotatebox{0}{\textbf{+}}} symbol.

The choice of cases to present in Figure~\ref{f11} was made to illustrate four points: 1) the weaker field cases are erratic, 2) the bisector fields show less dependence on position in the line than is the case for the Stokes $V$ field, 3) for the stronger field cases the slope dependence on spectral sampling is similar to the Stokes $V$ dependence on $\delta\lambda$ and 4) there does not seem to be a dependence of the results on the center-to-limb angle.  For field strengths above about 40 gauss, the patterns for the cases shown in the top four plots are typical.   For the weaker fields there is a considerable range of behavior with that for the bottom left case being an extreme.  The implied line core field for this case is only 20\%\ of that implied at $\Delta\lambda=84$m\AA.  Other cases of weak field actually have the core field greater than that at $\Delta\lambda=84$m\AA.  In contrast the higher field cases show the pattern of the bisector field being less dependent on $\Delta\lambda$ than the Stokes $V$ field is on $\delta\lambda$ for portions of the line nearer the line core.

\begin{figure}
\begin{center}
\parbox{3.8in}{
\resizebox{3.8in}{!}{\includegraphics{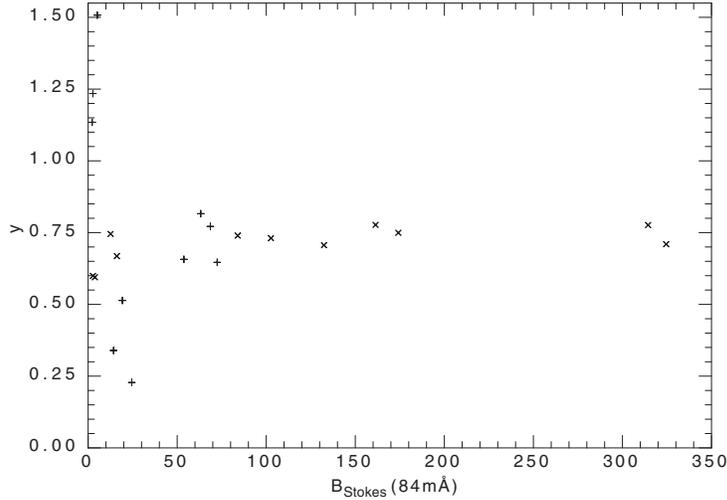}}
\caption{
This figure shows the factor $y$ which converts the slope $\eta^{5233,84}_{5250,39}$ into the correction factor that will yield our recommended magnetic field.  The \protect\raisebox{-2.5pt}{\protect\rotatebox{45}{\textbf{+}}} symbols represent line profiles observed with center-to-limb angles less the 45$^\circ$ while the \protect\raisebox{0pt}{\protect\rotatebox{0}{\textbf{+}}} symbols are for profiles with center-to-limb angles greater than 45$^\circ$. One point with $B=558$G and $y=0.689$ is not shown to avoid compressing the abscissa scale.  This point has a center-to-limb angle of 10$^\circ$. 
}
\label{f12}}
\end{center}

\end{figure}

\begin{figure}
\begin{center}
\parbox{3.8in}{
\resizebox{3.8in}{!}{\includegraphics{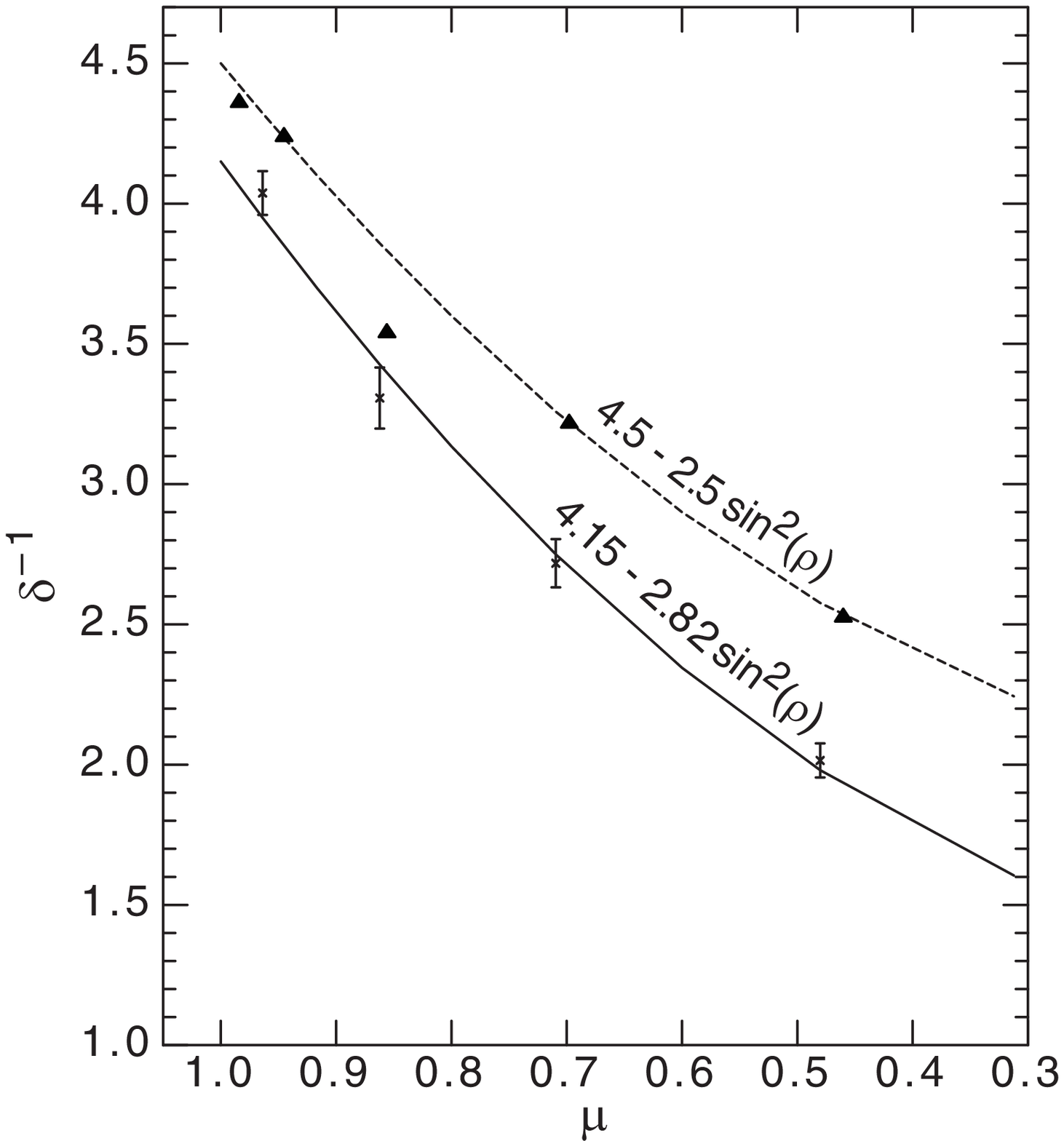}}
\caption{
This figure shows the final scale factor $\delta^{-1}$ 
which converts the observed magnetic field for $\lambda5250$\AA$\pm39$m\AA\ into the recommended magnetic field strength that gives our best estimate of the magnetic flux from each pixel emerging into the region above the solar atmosphere.  The \protect\raisebox{-2.5pt}{\protect\rotatebox{45}{\textbf{+}}} symbols with the error bars give the values of $\eta^{5250,39}_{5233,84}$ with their formal uncertainties multiplied by the average value of $y$ of 0.73. The solid line is the fit to $\sin^2(\rho)$ with the fitting coefficients given above the line.  In a similar fashion the filled triangles give the points from \protect\inlinecite{1992ASPC...26..265U} 
with dashed line showing the fit to these points and the fitting coefficients in use by \protect\inlinecite{1995ApJ...447L.143W}.
}
\label{f13}}
\end{center}
\end{figure}

Since the Stokes $V$ field $B_{\rm Stokes}$ gives a good reproduction of the correlation coefficients from the scatter diagrams and since the core parts of the bisector field $B_{\rm Bi}(\Delta\lambda)$ are less sensitive to $\Delta\lambda$ than $B_{\rm Stokes}(\delta\lambda)$ is to $\delta\lambda$, we adopt a model that the most appropriate field to use in determining the magnetic flux above the solar atmosphere is $B_{\rm Bi}(\Delta\lambda)$ where $\Delta\lambda$ approaches zero.  We believe that correlation coefficient field as well as the Stokes $V$ field near the line core are influenced by extra line broadening in the magnetized flux tube.  Since we are unable to observe this line profile we must use the profile from a standard quiet region which has a steeper slope near the line core so that the apparent falloff in field strength is due to division by too large a number.  It is possible to estimate some properties of the spectral line profile in the magnetized flux tube by integrating the Stokes $V$ parameter.  This gives a shape near the line core which is indeed broader compared to that of the unmagnetized line in the case of the observations of July 13, 2007.  The bisector field does not depend on the line slope but only on the requirement of equal intensity on the red and blue wings so that its falloff toward the line core must be a consequence of some other effect such as flux cancelation.  To combine the results of the magnetogram scatter diagrams and the line profile data using our adopted model, we identify the line profile quantity $B_{\rm Stokes}(84\hbox{m\AA})$ as corresponding to $B_{5233,84}$ used in the scatter diagram and we identify the line profile quantity $B_{\rm Bi}(29\hbox{m\AA})$ as the recommended field.  The ratio:
$$y={B_{\rm Bi}(29\hbox{m\AA})\over B_{\rm Stokes}(84\hbox{m\AA})}$$
is then a factor which will convert $\eta^{5233,84}_{5250,39}$ into the final correction factor for our observed magnetograms using $\lambda5250\hbox{\AA}\pm39\hbox{m\AA}$ in the MWO synoptic program.  We show the derived values of $y$ in Figure \ref{f12} as a function of $B_{\rm Stokes}(84\hbox{m\AA})$.

It is noteworthy that for areas with fields $B_{\rm Stokes}(84\hbox{m\AA})$ larger than about 30 gauss the values of $y$ are quite stable and do not depend on the center-to-limb angle.  The variations in $y$ for smaller fields are not due to error of measurement of the fields but rather are a consequence of some structural property of the solar surface.  For our purpose, we do not need to understand the cause for this behavior and can simply adopt the value of $y$ to be $0.73\pm0.015$ which is the average and error of the mean of all the cases with $B_{\rm Stokes}(84\hbox{m\AA})>30\;$gauss.
Examination of Figure \ref{f5a} reveals that the raw $V$ profiles for the weaker field cases are much less regular than the stronger field cases.  The flux tube structure implies that the weaker field areas have a small number of flux tubes.  The $V$ irregularities may be a consequence of statistical fluctuations around a fairly stable mean structure.  Theoretical studies of the $\lambda5233$\AA\ line would be helpful in understanding the properties of this line in the magnetized and unmagnetized solar atmosphere as well as providing a separate indication of the relationship between the various observed magnetic field strength indicators and the magnetic flux emerging into the heliosphere.

Based on the adoption of the $B_{\rm Bi}(29\hbox{m\AA})$ as the recommended field value, we can use the magnetogram correlation diagrams to obtain a new scale factor for the Mt.\ Wilson 150-foot tower telescope synoptic program which utilizes the FeI line at $\lambda5250$\AA.  For notation, we retain the traditional quantity $\delta^{-1}$ first defined by \inlinecite{1972SoPh...22..402H} wherein the recommended field $B_{\rm Rec}$ is:
$$B_{\rm Rec} = \delta^{-1} B_{5250,39}\ .$$
The previous correlation analysis by \inlinecite{1992ASPC...26..265U} as represented by \inlinecite{1995ApJ...447L.143W} was:
$$\delta^{-1}=4.5-2.5\sin^2(\rho)\ .$$ 
The present analysis is shown in Figure \ref{f13} and gives:
$$\delta^{-1}= y\eta^{5233,84}_{5250,39}=4.15-2.82\sin^2(\rho)\ .$$
Note that this correction factor is simply to take into acount line formation processes and does not include effects related to the geometry of the field.  In particular an additional $\sec(\rho)$ correction is required if one adopts a model wherein the magnetic field is radial.

\section{Discussion}
\label{disc}

Our principal result is given above and provides a small correction to the procedure previously applied to the Mt.\ Wilson Observatory magnetogram observations.  The previous description of this correction factor as a saturation correction is false.  The various interpretations on the polarimetry of the spectral lines used in our observations differ from one another due to processes involved with transfer of radiation through an inhomogeneous, magnetized atmosphere.  A full analysis of this problem is beyond the scope of the present investigation but is obviously needed to fully understand the changes in shape of the polarized spectral lines.  We base our recommended correction factor on the fact that the magnetic field deduced from the spectral line bisector is less variable than the field deduced from the monochromatic Stokes $V$ parameter.  We further adopt the magnetic field implied by the properties of the spectral line near its core on the grounds that this part of the line is closest to the outer edge of the solar atmosphere.  We find that the line profile properties are most stable in regions where the field strength is significantly larger than that found in the completely quiet sun.  The line profiles for the weakly magnetized regions are well determined by our system but they show substantial deviations from symmetry.  We suspect that this is a consequence of the reduced effect of adjacent flux tubes as a constraint on the flux tube structure in cases where the field is so weak that there are no adjacent flux tubes.  The unconstrained flux tubes can then adopt a wider range of structure including perhaps cases where the field lines return locally to the solar interior rather than extending out of the solar atmosphere.  The fact that the error of measurement of the fields is well below 1 gauss for our line profile system means that the measured profiles can be used to constrain models of flux tubes in regions where the field strength is very low.

Our determination that the differing deduced magnetic field strength for differing spectral sampling configurations has a general potential impact on the measurement of magnetic fields with other observing systems.  All programs provide maps which give a single magnetic field strength for each observed point.  Had the adjustment of the Mt.\ Wilson fields been a consequence of a saturation effect unique to this system, the potential implications would also have been limited to the Mt.\ Wilson system.  However, our determination that the effect comes from a spectral transfer effect means that other observing programs need to determine how to relate their field measurements to measurements made by other systems.  Even within a single observing programs, multiple deduced magnetic fields could occur depending on the exact specification of the algorithm bringing the spectropolarimetric data to a deduced magnetic field strength.  Examples where a direct comparison has been made on a pixel by pixel basis include \inlinecite{2005ApJS..156..295T} and \inlinecite{2008SoPh..tmp..125D}.  We are also in a position to update discussion from \ref{profandmag} which compares our scale factor to the deductions of \inlinecite{2003SoPh..213..213B}.  Those authors concluded from Advanced Stokes Polarimeter data that the MDI magnetic fields are $0.64\pm0.013$ times the correct value.  After combining the factor from \inlinecite{2005ApJS..156..295T} to the new corrected $\lambda5233$\AA\ scale we find that the MDI magnetic fields are $0.619\pm0.018$ times the recommended value at the center-to-limb distance used by \inlinecite{2003SoPh..213..213B}.
This agreement within errors of measurement gives support to our approach in the calibration of the $\lambda5250$\AA\ magnetograms.

\begin{acks}
The authors thank J.\ Harvey and J.\ Todd Hoeksema for their comments on drafts of this paper.  This work has been supported by NASA, NSF and ONR through a series of grants.  Most recently support has come from the project NASA/HMI 16165880-26967-G at Stanford University and from NSF through grant NSF ATM-0517729.
\end{acks}



\end{article} 
\end{document}